\documentclass[prb, twocolumn, notitlepage, superscriptaddress]{revtex4-1}
\usepackage{amsmath,amsfonts, amssymb, amsthm, dsfont}
\usepackage{yfonts}
\usepackage{bm}
\usepackage{mathrsfs}
\usepackage{graphicx}
\usepackage{verbatim}
\usepackage{hyperref}
\usepackage{wasysym}
\usepackage{tikz}
	\usetikzlibrary{calc}

\renewcommand{\vec}[1]{{\mathbf #1}}
\newcommand{\ket}[1]{|#1\rangle}

\renewcommand{\vr}{{\vec{r}}}

\newcommand{\comments}[1]{}
\newcommand{\mb}[1]{\mathbf{#1}}

\newcommand{\Ref}[1]{ Ref.~[\onlinecite{#1}]}

\newcommand{\U}{\ensuremath{\mathsf{U}(1)}}
\def\H{\mathcal{H}}
\def\Z{\mathbb{Z}}

\makeatletter
\def\l@subsubsection#1#2{}
\makeatother

\usetikzlibrary{decorations.pathreplacing,decorations.markings}
\tikzset{middlearrow/.style={
        decoration={markings,
            mark= at position 0.55 with {\arrow{#1}} ,
        },
        postaction={decorate}
    }
}
\begin{document}

\title{Fractonic Matter in Symmetry-Enriched $\mathrm{U}(1)$ Gauge Theory}
\author{Dominic J. Williamson}
\affiliation{Department of Physics, Yale University, New Haven, CT 06511-8499, USA}
\author{Zhen Bi}
\affiliation{Department of Physics, Massachusetts Institute of Technology, Cambridge, MA 02139, USA}
\author{Meng Cheng}
\affiliation{Department of Physics, Yale University, New Haven, CT 06511-8499, USA}

\begin{abstract}
In this work we explore the interplay between global symmetry and the mobility of quasiparticle excitations. We show that fractonic matter naturally appears in a three dimensional $\U$ gauge theory, enriched by global $\U$ and translational symmetries, via the mechanism of anyonic spin-orbital coupling.  We develop a systematic understanding of such symmetry-enforced mobility restrictions in terms of the classification of $\U$ gauge theories enriched by $\U$ and translational symmetries. We provide a unified construction of these phases by gauging layered symmetry-protected topological phases.
\end{abstract}

\maketitle

Recently, a new kind of topological quantum phenomenon has been discovered in three-dimensions, namely emergent particle excitations with restricted mobility. A variety of exactly-solvable lattice models have been constructed that exhibit completely \emph{immobile} particle excitations~\cite{Chamon2005,CastelnovoPRB2010, BravyiAOP2011, CastelnovoPM2012, Haah, VijayPRB2015, VijayPRB2016, YoshidaPRB2013}, called ``fractons''. These include the so-called type-I fracton models~\cite{VijayPRB2015, VijayPRB2016}, exemplified by the X-cube model, in which composites of fractons become subdimensional particles, and the more exotic type-II fracton models, such as Haah's cubic codes~\cite{Haah}, where all particle excitations are immobile~\cite{BravyiPRL2011,bravyi2013quantum,kim20123d,haah2014bifurcation,KimPRL2016}. Many further generalizations are being pursued, such as non-Abelian fracton phases~\cite{VijayFu2017, Prem2018, Song2018}, and ``twisted'' fracton models~\cite{YYZ2018, Song2018}. Various aspects of gapped fracton phases are being actively investigated~\cite{Haah2013,haah2013commuting,haah2016algebraic,Vijay2017,MaPRB2017,MaPRB2018, HePRB2018, ShiPRB2018, PremPRB2017, SlaglePRB2017, SlaglePRB2018, Shirley2017, Shirley2018,shirley2018Fractional,shirley2018Foliated, SchmitzPRB2018,PhysRevB.97.155111, DevakulPRB2018, HsiehPRB2017,HalaszPRL2017}.  On the other hand, it was also found that a large class of gapless phases, whose low-energy theory consists of higher-rank gauge fields, also support matter fields with restricted mobility~\cite{XuPRB2006, Rasmussen2016, PretkoPRB2017a, PretkoPRB2017b, PretkoPRB2017c, PretkoPRL2018,PremPRB2018,Pretko2018,PretkoWitten,ma2018higher,PretkoPRD2017, BulmashPRB2018,Gromov2017,Slagle2018, bulmash2018generalized,KumarArxiv2018}, and can be connected to gapped fracton phases via Higgs transitions~\cite{BulmashPRB2018, Ma2018}. As a genuinely new class of emergent quantum order, fractons have significantly broadened the horizon of 3D quantum phases~\cite{nandkishore2018fractons}.

In this work we consider how global symmetries, together with translation symmetry, affect the mobility of quasiparticle excitations.  It is known that fracton phases can emerge in translationally invariant systems with  \emph{subsystem} symmetries~\cite{ VijayPRB2016,WilliamsonPRB2016,kubica2018ungauging,ShirleyGauging2018, Devakul2018b,you2018subsystem,devakul2018universal,subsystemphaserel,Williamson2018, NussinovAP2009}:  if the symmetry acts nontrivially only on a lower-dimensional subsystem, then moving the charged particles out of the submanifold is clearly forbidden by the symmetry.  Here instead we consider the interplay between the translational symmetry and certain ordinary \emph{global internal} symmetries (\emph{i.e.} ``0-form'' symmetries~\cite{Gaiotto2015}).  
In particular, we find scenarios where the global symmetry quantum numbers of excited quasiparticles depend on their positions in a nontrivial way. As a result moving them requires operators that are charged under the symmetry. Therefore in the presence of these global symmetries, the mobility of the quasiparticles is restricted. If the global symmetry is then gauged, the restricted particles become fractons since charged operators are not allowed in the gauge-invariant Hilbert space. We find a natural realization of quasiparticles with such symmetry-enforced restricted mobility in a $\U$ spin liquid phase enriched by a global $\U$ symmetry, where the global symmetry effectively imposes electric dipole conservation~\cite{ShirleyGauging2018,Pretko2018,PretkoGauge2018,KumarArxiv2018}. 

More generally we reveal the relation between symmetry restrictions on the mobility of quasiparticles and symmetry-enriched topological orders, where the actions of translation and global symmetries on quasiparticle excitations do not commute. This line of thinking proves fruitful as we can systematically classify $\U$ gauge theories enriched by translation and certain global symmetry, and identify new examples of fractonic matter in these theories. We propose that all such theories can be constructed by gauging layered symmetry-protected topological (SPT) phases.

\section{Translation symmetry fractionalization in gapped topological phases}
\label{sec:Zn}
We start from a family of toy examples in two dimensions, which do not exhibit true fractonic behaviors but form close analogs.  Consider a translation-invariant system with an internal symmetry group $G$. We will assume that the system is in a symmetric topologically ordered phase. Based on the general formalism in \Ref{SET} (see also \Ref{Chen2014} and \Ref{Tarantino_SET}), \Ref{ChengPRX2016} classified the nontrivial actions of the symmetries on the quasiparticle excitations, \emph{e.g.} anyons in the gapped 2D topological order. First of all, translation symmetries may permute anyon types. This is an interesting symmetry action which we return to later. For now, we focus on the cases where anyons transform projectively under the symmetries. Of particular interest to us is the so-called ``\emph{anyonic spin-orbit coupling}'' (ASOC), which refers to the nontrivial interplay between translation and internal symmetry quantum number. Simply speaking, as an anyon is transported along some path in space, the $G$ symmetry charge of the whole system is changed accordingly. (Here the $G$ symmetry charge refers to a one-dimensional representation of $G$.) In other words, the string operators that move anyons are ``charged'' under the symmetry $G$. Mathematically such ASOC can be rigorously defined for translations along certain direction together with the internal symmetry $G$~\cite{ChengPRX2016}, and it can be classified~\cite{ChengPRX2016} by the second cohomology class $\H^2[\Z\times G, \mathcal{A}]=\H^1[G, \mathcal{A}] \times \H^2[G, \mathcal{A}]$, where $\mathcal{A}$ is the group of Abelian anyons in the symmetric topological order. Here $\H^2[G, \mathcal{A}]$ corresponds to the fractionalization of the $G$ symmetry itself, and $\H^1[G, \mathcal{A}]$ describes ASOC.

For a concrete example, let us consider the $\mathbb{Z}_N$ toric code model on a square lattice~\cite{qdouble}. The Hamiltonian is given by
\begin{equation}
{H}=-\sum_\vr (B_\vr+B_\vr^\dag) - \sum_\vr (A_\vr+A_\vr^\dag),
	\label{}
\end{equation}
where the plaquette terms $B_\vr$ and vertex terms $A_\vr$ are defined as
\begin{align}
B_\vr=\begin{tikzpicture}[scale=0.50, baseline={([yshift=-.5ex]current  bounding  box.center)}]
\draw (1,1) -- (-1,1);
\draw (-1,1) -- (-1,-1);
\draw (-1,-1) -- (1,-1);
\draw (1,-1) -- (1,1);
\draw (0,1.4) node {$Z^\dag$};
\draw (0,-1.4) node {$Z$};
\draw (-1.45,0) node {$Z^\dag$};
\draw (1.45,0) node {$Z$};
\draw (-1.3,-1.3) node {$\vr$};
\end{tikzpicture}\, ,
&&
A_\vr = \begin{tikzpicture}[scale=0.50, baseline={([yshift=-.5ex]current  bounding  box.center)}]
\draw (0,-1) -- (0,1);
\draw (-1,0) -- (1,0);
\draw (0,1.4) node {$X$};
\draw (0,-1.4) node {$X^\dag$};
\draw (-1.5,0) node {$X^\dag$};
\draw (1.4,0) node {$X$};
\draw (0.3,0.35) node {$\vr$};
\end{tikzpicture}\, .
\end{align}
Here $Z$ and $X$ are $\mathbb{Z}_N$ clock and shift operators acting on $\mathbb{Z}_N$ spins on the edges, which satisfy $ZX=e^{\frac{2\pi i}{N}}XZ$. It is useful to view the $\Z_N$ toric code as a $\Z_N$ lattice gauge theory, where $A_\vr$ terms energetically enforce the $\Z_N$ Gauss's law.

One can identify two types of excitations of the Hamiltonian: the electric (magnetic) excitations corresponding to violations of vertex (plaquette) terms. The elementary electric (magnetic) particle is denoted by $e$ ($m$). $e$ excitations can be created and moved around by a string of $Z$ or $Z^\dag$ along a certain path on the lattice. We adopt the convention that a vertex violation with $A_\vr$ eigenvalue $e^{2\pi i /N}$ is defined as the elementary $e$ excitation. For instance, the following string operator
\begin{equation}
	W=\prod_{x_L\leq x< x_R}Z_{(x,y),\hat{x}}^\dag 
	\, ,
	\label{}
\end{equation}
creates an $e-\bar{e}$ pair at the two endpoints $(x_L,y)$ and $(x_R,y)$, where the subscript $\{\vr,\hat{i}\}$ denotes the edge starting from site $\vr=(x,y)$ along the $\hat{i}$ axis.

Now we consider two global $\Z_N$ symmetries generated by the following operators,
\begin{align}
	S_{\hat{x}} = \prod_\vr X_{\vr,\hat{x}}\, , && S_{\hat{y}} = \prod_\vr X_{\vr,\hat{y}}\, .
	\label{eqn:SinZN}
\end{align}
The toric code Hamiltonian respects both of them. The two-anyon state $W\ket{0}$ has a nonzero $S_{\hat{x}}$ quantum number:
\begin{equation}
	S_{\hat{x}} W\ket{0} = e^{\frac{2\pi i}{N}(x_R-x_L)}W\ket{0}\, .
	\label{}
\end{equation}
In a sense, $S_{\hat{x}}$ measures the $\Z_N$ dipole moment along the $\hat{x}$ direction. Therefore, motions of the $e$ anyon along $\hat{x}$ change the $S_{\hat{x}}$ eigenvalue of the state. Consider moving an $e$ anyon by $l$ units along $\hat{x}$, whenever $l$ is not a multiple of $N$ the process is not allowed if the $S_{\hat{x}}$ symmetry is preserved. Thus the symmetry demands that an $e$ anyon can only move in steps of length $N$ along $\hat{x}$. Similar discussion applies for the $S_{\hat{y}}$ symmetry. On the contrary, the mobility of the $m$ excitations is not affected by these symmetries. 

\begin{figure}[ht]
	\centering
	\includegraphics[width=\columnwidth]{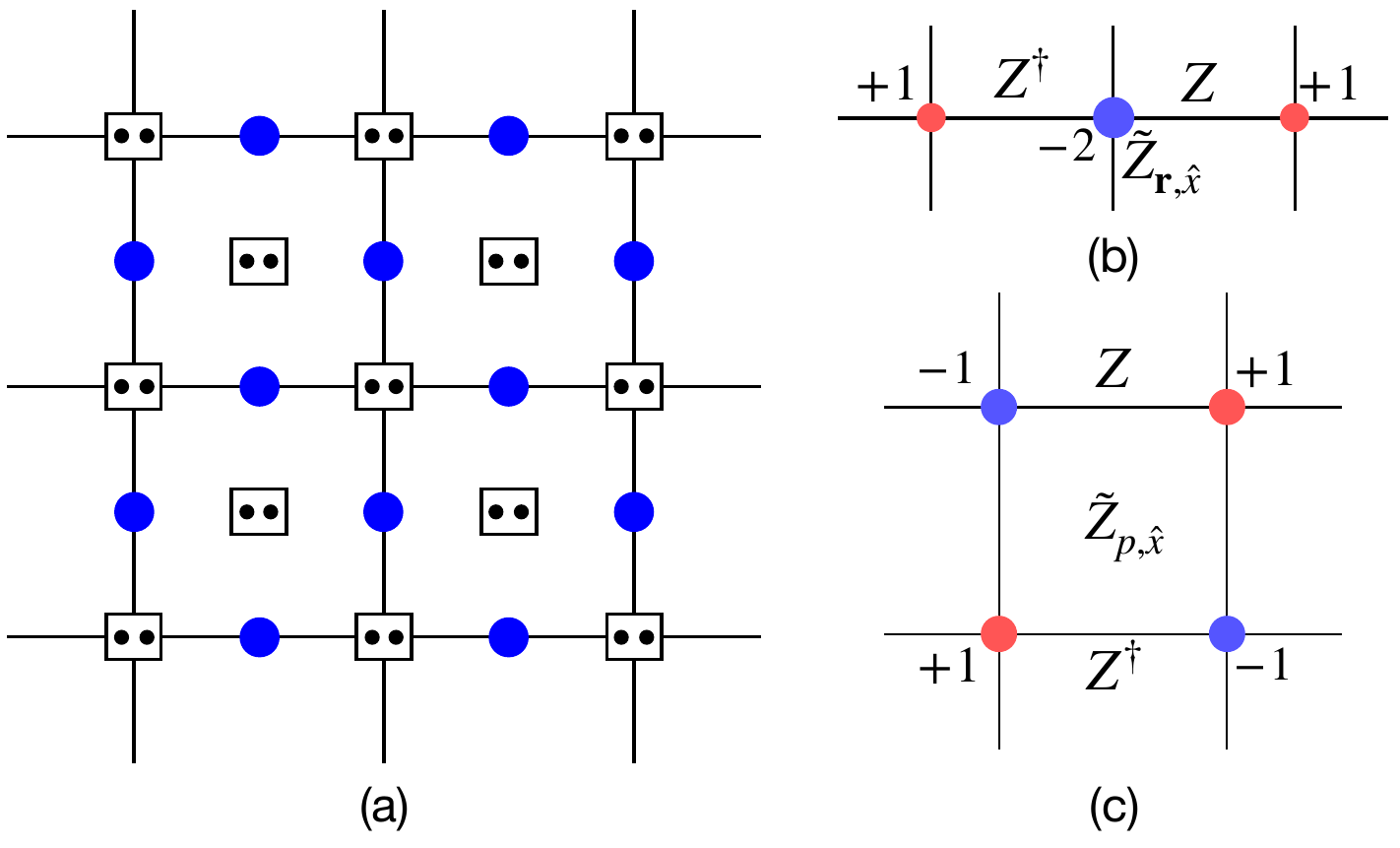}
	\caption{(a) Gauging $S_{\hat{x}}$ and $S_{\hat{y}}$ symmetry in a $\Z_N$ toric code model. Filled circles are original spins. Squares are gauge field spins, inside which the two dots represent gauge fields for $S_{\hat{x}}$ and $S_{\hat{y}}$ symmetries, respectively. (b) $e$ charge configuration created by Eq. \eqref{eqn:Zsite}. (c) $e$ charge configuration created by Eq. \eqref{eqn:Zplaq}. In (b) and (c) the numbers denote the corresponding $\Z_N$ gauge charge. }
	\label{fig:lattice2d}
\end{figure}

 Since $S_{\hat{x}}$ and $S_{\hat{y}}$ are unitary symmetries, we can promote them to gauge symmetries~\cite{wegner1971duality,Gaugingpaper,williamson2014matrix}. After introducing the dynamical gauge fields, only gauge-invariant operators are physical. Let us write down the gauged model explicitly. We introduce additional $\Z_N$ spins that serve as the gauge fields. We remark that only the spins on the $\vr,\hat{x}$ links are acted upon by $S_{\hat{x}}$. These spins also form a square lattice. The new ``gauge'' spins live on bonds of this square lattice, which correspond to the sites and the plaquette centers of the original square lattice. They are denoted by $\tilde{X}/\tilde{Z}_{\vr,\hat{i}}$ and $\tilde{X}/\tilde{Z}_{p,\hat{i}}$, where $\vr$ stands for the sites, $p$ stands for the plaquettes and $\hat{i}=\hat{x},\hat{y}$. We label a plaquette by the coordinate of its center. The lattice geometry is illustrated in Fig. \ref{fig:lattice2d} (a).

 Since the original spins are sources of gauge fields, we impose Gauss's law constraints:
 \begin{equation}
	 \begin{gathered}
		 X_{\vr,\hat{x}}\tilde{X}_{\vr,\hat{x}}^\dag\tilde{X}_{\vr+\mb{\hat{x}},\hat{x}}\tilde{X}_{\vr+\frac{\mb{\hat{x}}}{2}-\frac{\mb{\hat{y}}}{2},\hat{x}}^\dag\tilde{X}_{\vr+\frac{\mb{\hat{x}}}{2}+\frac{\mb{\hat{y}}}{2}, \hat{x}}=1\\
		 X_{\vr,\hat{y}}\tilde{X}_{\vr,\hat{y}}^\dag\tilde{X}_{\vr+\mb{\hat{y}},\hat{y}}\tilde{X}_{\vr-\frac{\mb{\hat{x}}}{2}+\frac{\mb{\hat{y}}}{2},\hat{y}}^\dag\tilde{X}_{\vr+\frac{\mb{\hat{y}}}{2}+\frac{\mb{\hat{x}}}{2}, \hat{y}}=1
	 \end{gathered}
	 \label{}
 \end{equation}
 for all $\vr$.

We also have to modify the plaquette term by the minimal coupling:
\begin{equation}
	\tilde{B}_p= B_p\tilde{Z}_{p,\hat{x}}\tilde{Z}_{p,\hat{y}}^\dag
	\label{}
\end{equation}

The last step is to add plaquette interactions for the new gauge fields:
\begin{equation}
	-K'\sum_\vr \tilde{Z}_{\vr,\hat{i}}\tilde{Z}_{\vr+\frac{\mb{\hat{x}}}{2}+\frac{\mb{\hat{y}}}{2},\hat{i}}\tilde{Z}_{\vr+\hat{y},\hat{i}}^\dag \tilde{Z}_{\vr-\frac{\mb{\hat{x}}}{2}+\frac{\mb{\hat{y}}}{2},\hat{i}}^\dag +\text{h.c.}.
	\label{}
\end{equation}

To see the mobility of excitations, let us construct gauge-invariant operators. Clearly, we may still apply strings of $X$ operators to create/move the $m$ (plaquette) excitations, and similarly for $\tilde{X}$. For strings composed by $Z$'s or $\tilde{Z}$'s, Gauss's law constraints imply that they must be combined in certain ways. The basic building blocks are
\begin{equation}
	\tilde{Z}_{\vr,\hat{x}}Z_{\vr-\mb{\hat{x}},\hat{x}}^\dag Z_{\vr,\hat{x}},
	\label{eqn:Zsite}
\end{equation}
and
\begin{equation}
	\tilde{Z}_{\vr+\frac{\mb{\hat{x}}}{2}+\frac{\mb{\hat{y}}}{2},\hat{x}}Z_{\vr+\mb{\hat{y}},\hat{x}} Z_{\vr,\hat{x}}^\dag
	\label{eqn:Zplaq}
\end{equation}
Both do not commute with nearby $A_\vr$ operators and create clusters of $e$ charges. The patterns of $e$ charge configuration are illustrated in Figs. \ref{fig:lattice2d} (b) and (c). They agree with the charge configuration pattern in the $(1,1)$ scalar charge theory discussed in \Ref{BulmashPRB2018}. In fact, one can show explicitly that if projected to the subspace defined by $\tilde{B}_p=1$, the gauged model can be mapped exactly to the Higgsed tensor gauge theory with the $(1,1)$ scalar charge~\cite{BulmashPRB2018}.

In the language of anyon models, we have essentially created three copies of $\Z_N$ toric code: the original one generated by $e$ and $m$, and a new one for each of the gauged symmetries, generated by $e_x$ and $m_x$ for $S_{\hat{x}}$ gauge symmetry, and $e_y$ and $m_y$ for $S_{\hat{y}}$. Under $T_{\hat{x}}$ translation we have:
\begin{equation}
	T_{\hat{x}}: e\rightarrow ee_x, m_x\rightarrow m_x\bar{m} \, ,
	\label{}
\end{equation}
where $\bar{m}$ is the antiparticle to $m$. 
To see the transformation of $m_x$, notice that $m_x$ is created by a dual string of $\tilde{X}$. However it does not commute with the $K$ term unless one attaches additional $X$ operators. Since the action of $T_{\hat{x}}$ permutes the $e$ particle's topological charge, there exists no local operator that can move $e$ by one step along $\hat{x}$. Of course, one can move an $e$ anyon by $N$ steps since $e$ is invariant under $T_{\hat{x}}^N$. A similar analysis applies for $T_{\hat{y}}$. Thus there are no true fractons in this model. We also remark that the mobility of $m$ and $e'$ excitations are not affected.  

\section{Fractonic $\U$ gauge theory}
\label{sec:U1gauge}
In this section we focus on fractonic matter in 3D lattice $\U$ gauge theories. We first present an example of symmetry-enforced fractonic matter, and then discuss its possible realizations in models of $\U$ spin liquids. In the end, we present a general classification of fractonic $\U$ gauge theories.

\subsection{An example of fractonic global $\U$ symmetry}
\label{sec:example}
In the above $\Z_N$ gauge theory example, due to the global symmetry involved in the ASOC being finite, excitations could always move in some special long steps. It is then natural to expect that a global $\U$ symmetry with nontrivial ASOC could prevent quasiparticle excitations from moving completely. However, for particle excitations in a gapped phase, this is impossible because $\H^1[\U,\mathcal{A}]=\Z_1$, for any finite Abelian group $\mathcal{A}$. Instead we consider a similar effect in a gapless $\U$ gauge theory (in a sense this is the $N\rightarrow\infty$ limit of the $\Z_N$ model).

First we review the $\U$ lattice gauge theory. Consider a $d$-dimensional cubic lattice, with one rotor on each bond, which is described by a pair of conjugate variables ${{A}_{\vr\vr'}=-{A}_{\vr'\vr}}, {{E}_{\vr\vr'}=-{E}_{\vr'\vr}}$. They satisfy the canonical commutation relation 
	$[{A}_{\vr\vr'}, {E}_{\vr\vr'}]=i$,
and commute on different bonds. Here the ${A}$'s are $2\pi$-periodic. In other words the $\U$ gauge field is compact. ${E}$'s take integer eigenvalues. We will denote ${A}_{\vr,\hat{\mu}}\equiv {A}_{\vr,\vr+\hat{\mu}}$, and similarly ${E}_{\vr,\hat{\mu}}\equiv {E}_{\vr,\vr+\hat{\mu}}$.

The Hamiltonian of the gauge theory is given by
\begin{equation}
	H=-K\sum_p \cos (\nabla\times \mb{A}) + \frac{U}{2}\sum_{e}E_{e}^2 + \Delta\sum_\vr (\nabla\cdot\mb{E})_\vr^2
	\, ,
	\label{eqn:U1gauge}
\end{equation}
where ${(\nabla \cdot \mb{E})_\vr = \sum_{\vr'\in \text{NN}(\vr)}E_{\vr\vr'}}$ is the lattice divergence, and NN stands for nearest neighbors. ${(\nabla\times \mb{A})_p = \sum_{e\in p} \epsilon_p^e A_{e}}$ denotes the lattice curl for a plaquette $p$, with $\epsilon_p^e$ being $+1$ if the orientation of edge $e$ matches that of the boundary of plaquette $p$, and $-1$ otherwise.

We remark that the last term in the Hamiltonian essentially imposes Gauss's law as an energetic constraint. Because it commutes with all other terms, we can divide the Hilbert space into sectors labeled by different configurations of charges:
\begin{equation}
	(\nabla \cdot \mb{E})_\vr = q_\vr,\: q_\vr\in\Z
	\, .
	\label{}
\end{equation}

It is well known that a compact $\U$ gauge theory has a Coulomb phase in $d\geq 3$ for $U/K<(U/K)_\text{c}$~\cite{KogutRMP}, with the critical value $(U/K)_\text{c}\approx 1$. In the Coulomb phase, we can effectively ignore the compactness of the gauge field and take a continuum limit to obtain Maxwell electrodynamics. The low-energy excitations include propagating photons and gapped electric and magnetic charges. For $d=2$, a compact $\U$ gauge theory generally becomes confined due to the proliferation of monopoles~\cite{Polyakov}. Notice that in the ``fixed-point'' Hamiltonian of Eq.~\eqref{eqn:U1gauge}, all charges are static because every term commutes with the Gauss's law constraint. However under a generic perturbation the charges will become dynamical.

We now turn to global symmetries of the $\U$ gauge theory that exhibit an analog of the ASOC phenomenon in the $\Z_N$ toric code model.  If we view the $\U$ gauge theory as a limit of $\Z_N$ gauge theory with $N\rightarrow \infty$, then the analog of the symmetries defined in Eq.~\eqref{eqn:SinZN} are $\U$ symmetries generated by the following conserved charges:
\begin{align}
	S_{\hat{\mu}} = \sum_{\vr}E_{\vr, \hat{\mu}}\, , &&  \mu= x, y, z ,
	\label{}
\end{align}
where the sum is over all lattice sites $\vr$. 
It is straightforward to check that the gauge theory Hamiltonian in Eq.~\eqref{eqn:U1gauge} indeed commutes with each $S_{\hat{\mu}}$.

Let us show that $-S_{\hat{\mu}}$ is the electric dipole moment  of the charge distribution projected onto the ${\hat{\mu}}$ axis ${\mb{P}_{\hat{\mu}}=\sum_\vr q_\vr  \,\vr \cdot {{\hat{\mu}}}} $. 
As an example, consider a pair of charges, $+q$ and $-q$, where $+q$ is at the origin $(0,0,0)$ and $-q$ is at $\vr=(x,y,z)$. One can compute $S_{\hat{\mu}}$ for an arbitrary configuration of electric field lines consistent with the Gauss's law, and it is easy to see that
\begin{equation}
	S_{\hat{\mu}} = qr_{\hat{\mu}}.
	\label{}
\end{equation}
Mathematically, we can simply observe the following relation~\footnote{We thank Kevin Slagle for pointing this out.}  
\begin{equation}
	- \sum_\vr \mb{E}_{\vr} = \sum_\vr \vr\, (\nabla\cdot \mb{E})_\vr = \sum_\vr \vr \, q_\vr.
	\label{}
\end{equation}
Therefore $- \mb{S}=\mb{P}$. As observed in \Ref{PretkoPRB2017a}, the conservation of electric dipole moment implies that no charges can move. Thus the mobility of electric charges becomes restricted if we impose the $S_{\hat{\mu}}$ symmetries.

The change of the total $S_{\hat{\mu}}$ quantum number caused by the motion of a charged particle along the $\hat{\mu}$ direction is the expected generalization of the ``anyonic spin-orbit coupling'' phenomenon in $\U$ gauge theory. In \Ref{ZouPRB2018a}, it was shown that $\U$ gauge theories enriched by a global symmetry group $G$ can be classified by projective representations of $G$ carried by electric and magnetic charges.  For simplicity, let us assume for now that $G$ is unitary and does not transform charge types. Then the projective representation on an electric/magnetic charge is classified by the second group cohomology $\H^2[G, \U]$. The ASOC requires both a global $\U$ symmetry (\emph{i.e.} one of the $S_{\hat{\mu}}$'s) and lattice translation symmetries. So we take $G=\Z\times \U$ where $\Z$ is one of the translations, and applying K\"unneth formula we find
	\begin{equation}
		\begin{split}
		\H^2[\U\times \Z, \U]
		&=\Z.
		\end{split}
		\label{}
	\end{equation}
The interpretation of $\Z$ is precisely the change of the global $\U$ charge as a particle is transported by one lattice unit.

To summarize, we have shown that electric charges in this model are in fact fractonic matter, if a global $\U$ symmetry that enforces conservation of dipole moments is present. 
We can gauge the global symmetry generated by $S_{\hat{\mu}}$ to make the electric charges truly immobile.  It is straightforward to write down lattice models following a similar procedure to that described in Sec.~\ref{sec:Zn}. It is natural to expect that the gauged model is closely related to tensor gauge theories with scalar charge~\cite{PretkoPRB2017a, PretkoPRB2017b}. In Sec.~\ref{sec:gauging} we explicitly carry out the gauging procedure and demonstrate that a higher rank tensor gauge theory with $(1,1)$ scalar charge indeed emerges in the gauged model~\cite{BulmashPRB2018}.

\subsection{Fractonic $\U$ spin liquid}
$\U$ electrodynamics can emerge as a low-energy effective theory of physical spin systems~\cite{HermelePRB2004}. For example, potential realizations of emergent electromagnetism were proposed in the so-called ``quantum spin ice'' state of rare-earth pyrochlore materials~\cite{SavaryPRL2012, RossPRX2011}. In this section, we extend the observations made in Sec. \ref{sec:U1gauge}-A to spin models which can give rise to an emergent $\U$ gauge theory\cite{HermelePRB2004}. We first consider the $\U$ spin liquid model on the cubic lattice, and then on the pyrochlore lattice.

\subsubsection{The cubic model}
Consider a model of spin-$1/2$'s on the cubic lattice, where the spins occupy the nearest-neighbor bonds and form a lattice of corner-sharing octahedra. Let us focus on the following Hamiltonian:
\begin{equation}
	H_\text{cub}=J_z\sum_\text{oct}(S^z_{\text{oct}})^2 - J\sum_{\Box} (S_1^+S_2^-S_3^+S_4^- + \text{h.c.}),
	\label{eqn:u1sl}
\end{equation}
where $S^z_\text{oct}=\sum_{\hat{\mu}} (S^z_{\vr, \hat{\mu}}+S^z_{\vr-\hat{\mu},\hat{\mu}})$. The numbering in the second term runs over the perimeter of each square plaquette.

To reveal the $\U$ gauge structure, following \Ref{HermelePRB2004} we first soften the constraint that $S=1/2$ on each site. We introduce rotor variables $n_{\vr\vr'}$ and $\phi_{\vr\vr'}$, where $\phi_{\vr\vr'}\in[0,2\pi)$, $n_{\vr\vr'}\in\mathbb{Z}$ and $[\phi_{\vr\mu},n_{\vr'\nu}]=i\delta_{\vr\vr'}\delta_{\mu\nu}$. We represent the spin variables as $S^z=n-1/2$ and $S^\pm=e^{\pm i\phi}$. For this mapping to be valid we must impose the hard-core constraint $n=0$ or $n=1$, which is achieved by including a repulsion $U\sum_{\vr\vr'}(n_{\vr\vr'}-1/2)^2$ in the Hamiltonian. We now further define link variables 
\begin{align}
	e_{\vr\vr'}=\epsilon_\vr n_{\vr\vr'}\, , && a_{\vr\vr'}=\epsilon_\vr \phi_{\vr\vr'}
	\label{eqn:defineEA}
\end{align}
Here we define $\epsilon_\vr=1$ or $-1$ when $\vr$ lies in the A or B sublattice. The Gauss's law constraint is given by
\begin{equation}
	(\nabla \cdot \mb{e})_\vr = \epsilon_\vr (S^z_\text{oct}+3).
	\label{}
\end{equation}
Notice that in the ground state $S^z_\text{oct}=0$, so there is actually a background of static charges $\pm 3$.  We then obtain the following Hamiltonian:
\begin{equation}
	H=\frac{U}{2}\sum_{\langle\vr\vr'\rangle}\left(e_{\vr\vr'}-\frac{\epsilon_\vr}{2}\right)^2 - K\sum_p \cos (\nabla\times\mb{a}).
	\label{}
\end{equation}

In the following we are interested in the Coulomb phase of the model in Eq. \eqref{eqn:u1sl}, which is known to exist for all values of $U$ from Monte Carlo simulations~\cite{BanerjeePRL2008, ShannonPRL2012}.

Let us now discuss the global symmetries in this system.  The spin model has an $\mathrm{O}(2)$ symmetry, generated by spin rotation around the $z$ axis and $\pi$ rotation around the $x$ axis. In particular, we find that
\begin{equation}
	\begin{split}
	\sum_{\vr,\hat{\mu}} S_{\vr,\hat{\mu}}^z &=\sum_{\vr,\hat{\mu}}\left(\epsilon_\vr e_{\vr,\hat{\mu}}-\frac{1}{2}\right)\\
	&= \sum_{\vr\in A,\hat{\mu}}e_{\vr,\hat{\mu}} - \sum_{\vr\in B,\hat{\mu}}e_{\vr,\hat{\mu}}+\text{const}\, .
	\end{split}
	\label{}
\end{equation}
The only constraint following from the conservation of total $S^z$ is that electric charges must move on the same sublattice, which is a well-known fact~\cite{HermelePRB2004}.

The fractonic symmetry we considered actually maps to the staggered magnetization:
\begin{equation}
	S_{\hat{\mu}}=\sum_{\vr}\epsilon_{\vr}S_{\vr,\hat{\mu}}^z = \sum_{\vr}e_{\vr,\hat{\mu}}.
	\label{eqn:sym-U1SL}
\end{equation}
Notice that there is a slight difference compared to the ``unfrustrated'' model, Eq. \eqref{eqn:U1gauge}, namely the ground state already has nonzero electric field lines emitting from sublattice A to sublattice B (e.g. $e_{\vr\vr'}=1$ for all $\vr\in A$ and $\vr'$ being the nearest neighbor B sites).
Again, one may show that $S_{\hat{\mu}}$ is completely determined by charge configurations, and the change of $S_{\hat{\mu}}$ is equal to the change of the electric dipole moment. Therefore the $\U$ gauge theory also exhibits ASOC and the conservation of $S_{\hat{\mu}}$ forbids motion of electric charges along $\hat{\mu}$. We emphasize that the ASOC phenomena, and consequently the symmetry-enforced fractonic behavior, are universal properties of the Coulomb phase of the $\U$ spin liquid model as long as the symmetry is preserved.

As pointed out in \Ref{HermelePRB2004}, the $J$ ring-exchange term can be generated from two-body spin exchange terms in the limit of large $J_z$. Examples of such terms consistent with the symmetry $S_{\hat{\mu}}$ in Eq.~\eqref{eqn:sym-U1SL} are 
\begin{equation}
	-\sum_{\vr,\vr'}\left(J_{\vr\vr',\hat{\mu}}^\pm S_{\vr,\hat{\mu}}^+ S_{\vr',\hat{\mu}}^- + \text{h.c.}\right),
	\label{}
\end{equation}
 where $\vr,\vr'$ belong to the same sublattice, 
 or ${ J'_{\vr\vr',\hat{\mu}}S_{\vr,\hat{\mu}}^+S_{\vr', \hat{\mu}}^+ + \text{h.c.}}$ where $\vr,\vr'$ belong to different sublattices.
While the latter choice breaks the global $S^z$ conservation, it still preserves the staggered magnetization symmetry $S_{\hat{\mu}}$.  The $\U$ spin liquid phase should extend to small but finite values of $J^\pm/J_z$. Since the gauge structure remains the same, we expect that the electric charges exhibit fractonic dynamics.

Alternatively, we may redefine the spin operators:
\begin{equation}
	S_{\vr,\hat{\mu}}^z\rightarrow \epsilon_{\vr} S_{\vr,\hat{\mu}}^z \, .
	\label{}
\end{equation}
which can be achieved by a unitary operator $U = \prod_{\vr\in B,\hat{\mu}}S^x_{\vr,\hat{\mu}}$. This transformation also sends
\begin{align}
	S^\pm_{\vr,\hat{\mu}}\rightarrow S^{\mp}_{\vr,\hat{\mu}}\, , \qquad \vr\in B\, .
	\label{}
\end{align}
Then the fractonic symmetry $S_{\hat{\mu}}$ becomes the total $S^z$ on edges along the $\hat{\mu}$-th direction.  A general Hamiltonian then takes the following form:
\begin{equation}
	\begin{split}
	H=&J_z\sum_{\vr} \Big(\sum_{\hat{\mu}} S_{\vr,{\hat{\mu}}}^z - \sum_{\hat{\nu}} S_{\vr-\hat{\nu},\hat{\nu}}^z\Big)^2 \\
	&- \sum_{\vr,\vr',\hat{\mu}}(J_{\vr\vr'}^{\hat{\mu}}S_{\vr,\hat{\mu}}^+ S_{\vr',\hat{\mu}}^- + \text{h.c.}) \, .
	\end{split}
	\label{}
\end{equation}
If we only preserve $\sum_{\hat{\mu}}S_{\hat{\mu}}$ which is the total $S^z$, we can allow almost any couplings $S^+S^-$.

We remark that these Hamiltonians can be effectively simulated by quantum Monte Carlo algorithms without a sign problem as long as the $J_{\vr\vr'}^{\hat{\mu}}$ coefficients are positive~\cite{HuangPRL2018}. It would be interesting to study the dynamics of fractonic charges in the model using such numerical simulations.

\subsubsection{The pyrochlore model}
Similar results can be derived for XXZ-type models of $\U$ spin liquid on the pyrochlore lattice. For the pyrochlore $\U$ spin liquid~\cite{HermelePRB2004}, spins reside on the sites of corner sharing tetrahedrons. The dual lattice of the pyrochlore structure is a diamond lattice. The A and B sublattice sites of the dual diamond lattice map to centers of the tetrahedrons with different orientations, and spins reside on the links of the dual diamond lattice. Let us consider the low-energy effective theory of the XXZ model on the pyrochlore lattice. The Hamiltonian is given by:
\begin{equation}
	H_\text{pyro}=J_z\sum_{\text{tetra}}(S^z_\text{tetra})^2-J\sum_{\hexagon}(S_1^+S_2^-S_3^+S_4^-S_5^+S_6^-+\text{h.c.}).
\label{eqn:pyro-model}
\end{equation} 
On each link, we can define $\U$ gauge variables in the same way as in Eq.~\eqref{eqn:defineEA} and when written in these variables the theory takes the same form as Eq.~\eqref{eqn:U1gauge} (without the charging energy $\Delta$ term). 

\begin{figure}[ht]
	\centering
	\includegraphics[width=0.8\columnwidth]{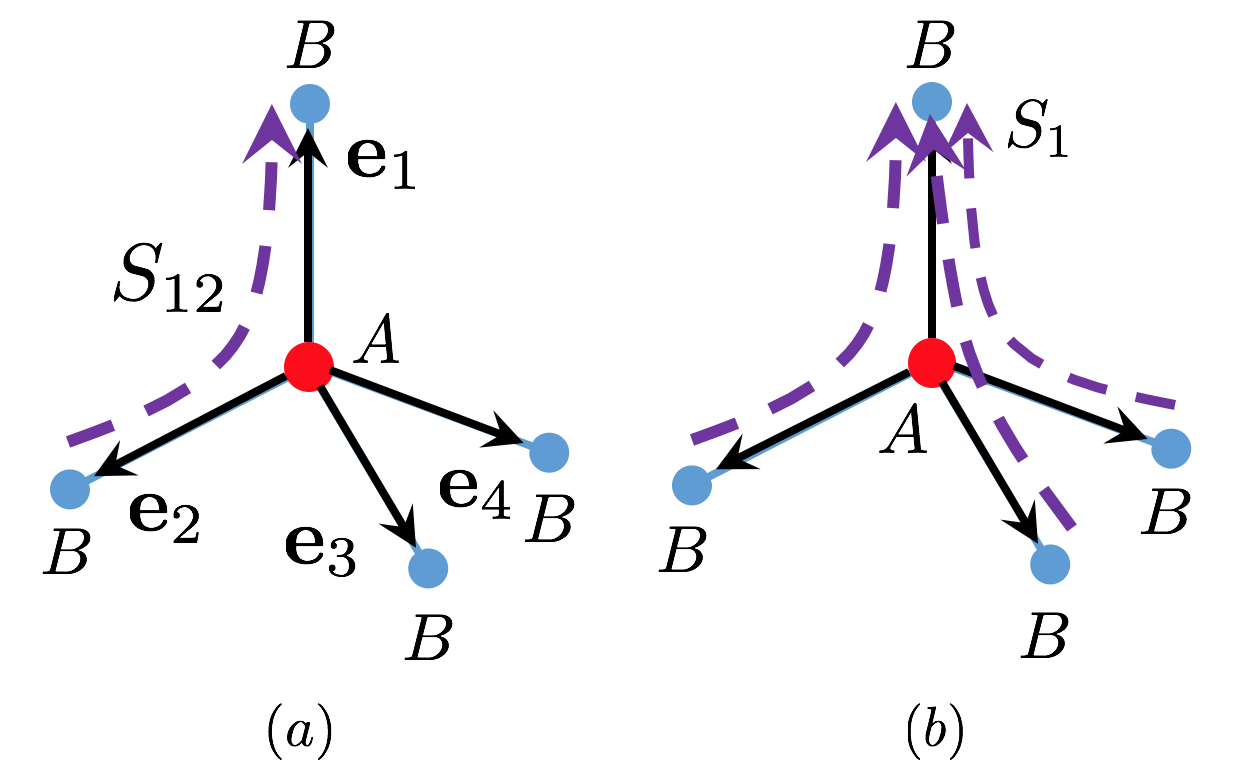}
	\caption{Examples of the definition of $S_{\mu\nu}$ and $S_{\mu}$. The unit cell of the diamond lattice is depicted in (a) and (b). The operator $S_{\mu\nu}$ measures the electric field lines that go through the $\mu\nu$ direction. (a) As an example $S_{12}$ is shown. (b) An example $S_\mu$ operator which measures the electric dipole moment of the system along the $\vec{e}_\mu$ direction. }
	\label{fig:diamond}
\end{figure}

We now introduce global symmetries similar to those given in Eq.~\eqref{eqn:sym-U1SL}, which constrain the mobility of electrically charged excitations in the pyrochlore model. The $S_{\hat{\mu}}$ operators defined in Eq.~\eqref{eqn:sym-U1SL}  can be thought of pictorially as collections of electric field lines along $\hat{\mu}$, which measure the total electric dipole moment along $\hat{\mu}$. The diamond lattice structure is more complicated than the cubic lattice. However, we can still find three independent ``fractonic'' symmetries in this model. A unit cell for the diamond lattice is shown in Fig.~\ref{fig:diamond}. We define six types of staggered magnetization operators:
\begin{equation}
S_{\mu\nu}=\sum_{\vec{r}\in A}\left(S^z_{\vec{r},\mu}-S^z_{\vec{r}+\vec{e}_\nu,-\nu}\right)=\sum_{\vec{r}\in A}e_{\vec{r},\mu}+\sum_{\vec{r}\in B}e_{\vec{r},-\nu},
\label{eqn:sym-pyro}
\end{equation}
where $\mu, \nu=1,2,3,4$ and $S_{\mu\nu}=-S_{\nu\mu}$. It is easy to check that $S_{\mu\nu}$'s commute with the Hamiltonian. Hence, they are symmetries of the $\U$ spin liquid. The $S_{\mu\nu}$ operator measures the number of electric field lines passing through the $\mu\nu$ direction. However, these symmetries are not independent. They satisfy the following constraints:
\begin{equation}
S_{\mu\nu}+S_{\nu\lambda}+S_{\lambda\mu}=0,
\end{equation}
where $\mu,\nu$ and $\lambda$ are all different. There are three independent constraints, so the number of independent symmetries is three, as expected.
 We can organize them in the following way,
\begin{equation}
S_{\mu}=\frac{1}{3}\sum_{\nu\neq\mu}S_{\mu\nu},
\end{equation}
where $\mu=1,2,3,4$ and $S_4$ is linear combination of $S_{1,2,3}$. $S_\mu$ measures total electric dipole moment along the $\vec{e}_\mu$ direction. 

If one of the $S_\mu$ symmetries is preserved, then the electric particle can only move in the plane normal to the $\vec{e}_\mu$ direction,  which is a honeycomb lattice. In addition, since the honeycomb lattice is rugged, the electric particle can only hop among the same sublattice sites. If we require two $S_\mu$ symmetries, then the electric particle can only hop along a line. If all three $S_\mu$'s are preserved in the system, the electric particle is immobile. Again, these symmetries do not put constraints on the mobility of the magnetic monopoles.

\subsection{General classification}
\label{sec:classification}
So far we have considered $\U$ gauge theory where the mobility of electric charges is restricted by a global $\U$ symmetry.  More generally, we can have particle excitations carrying both electric and magnetic charges, called dyons. In the previous model, Eq. \eqref{eqn:U1gauge}, because magnetic monopoles are created/moved by superpositions of $E_{\vr\vr'}$ operators, the global symmetries $S_{\hat{\mu}}$ do not have any effect on their mobility.  One natural question to ask: is it possible for both of them to have restricted mobility? What are the general constraints on the mobility of dyonic excitations beyond any specific models? Below we present partial results on a general classification of $\U$ gauge theories with fractonic matter, by relating the mobility of excitations to translation symmetry action.

\subsubsection{Symmetry-enforced fractonic matter}
Let us start from  $\U$ gauge theories with a global $\U$ symmetry and classify fractonic behavior of charged excitations enforced by the symmetry. As we have emphasized, the fractonic behavior of electric gauge charges can be understood in the framework of symmetry-enriched $\U$ gauge theories. We now consider the classification of $\U$ gauge theories enriched by $\U\times \Z^3$ symmetry. Since $\H^2[\U\times \Z^3, \U]=\Z^3$, we can characterize the symmetry-enriched $\U$ gauge theories by two integer vectors $(\mb{v}_e, \mb{v}_m)$, where $\mb{v}_e$/$\mb{v}_m$ labels the fractionalization class of electric/magnetic charges. To see the physical meaning, consider a generic dyonic excitation in the $\U$ gauge theory, labeled by the electric and magnetic charges $(q_e, q_m)$. As it is transported by $\vr$,  a global $\U$ charge 
\begin{equation}
	\vr\cdot(\mb{v}_e q_e + \mb{v}_m q_m)
	\label{}
\end{equation}
is acquired. Thus a $(q_e, q_m)$ dyon can only move in the plane perpendicular to the vector $q_e\mb{v}_e +q_m\mb{v}_m$.  We have also considered the case of multiple $\U$ symmetries in the Appendix~\ref{app:anomaly}.

The remaining question is whether there exists any 't Hooft anomaly associated with a given $(\mb{v}_e, \mb{v}_m)$ fractionalization class. If the anomaly does not vanish, it means that the corresponding fractionalization class (and hence the fractonic matter) can not actually be realized physically in three dimensions.  We can compute the 't Hooft anomaly explicitly using the formalism developed in \Ref{NingU1}, and details can be found in Appendix~\ref{app:anomaly}. There is no 't Hooft anomaly if and only if
\begin{equation}
	\mb{v}_e\times \mb{v}_m=0.
	\label{}
\end{equation}
Thus either one of $\mb{v}_e$ and $\mb{v}_m$ vanishes, or they must be parallel (or anti-parallel) vectors. In the latter case, we can find two co-prime integers $p,q$ such that $p\mb{v}_e+q\mb{v}_m=0$, and then a $(p,q)$ dyon is free to move. The lattice models we have studied so far realize $\mb{v}_e\neq 0, \mb{v}_m=0$. A physically distinct case is $\mb{v}_e=\pm \mb{v}_m$, where we can choose $p=\mp q =1$, and thus the free dyon is a fermion.

One may wonder how these other types of fractonic $\U$ gauge theories can be realized physically. We show below that they can be systematically constructed as gauged layered SPT phases.

\subsubsection{Intrinsically fractonic matter}
An important aspect of symmetry enrichment we have neglected thus far is the permutation of topological quasiparticle types by symmetries. For a $\U$ gauge theory, all universal properties of gapped quasiparticles are encoded in the charge lattice, and the intrinsic symmetries of such a lattice are the duality group $\mathcal{D}$. For $\U$ gauge theory with bosonic matter, $\mathcal{D}$ is a subgroup of the modular group SL$(2, \Z)$, generated by the following transformations:
\begin{equation}
	\begin{gathered}
	T: 
	\begin{pmatrix}
		q_e\\
		q_m
	\end{pmatrix}
	\rightarrow
	\begin{pmatrix}
		q_e+q_m\\
		q_m
	\end{pmatrix},\\	
	S:
	\begin{pmatrix}
		q_e\\
		q_m
	\end{pmatrix}
	\rightarrow
	\begin{pmatrix}
		-q_m\\
		q_e
	\end{pmatrix}.
	\end{gathered}
	\label{}
\end{equation}
Because $T$ does not preserve the exchange statistics of dyons, the actual duality group $\mathcal{D}$ is generated by $T^2$ and $S$. For $\U$ gauge theory with fermionic matter (i.e. the unit electric charge is a fermion), then the symmetry is the full duality group.

To describe the action of a global symmetry group $G$, we need to specify a group homomorphism $\rho: G\rightarrow \mathcal{D}$ from the global symmetry group $G$ to the duality group. In all previous discussions, we have assumed that $\rho$ maps $G$ to the identity element in $\mathcal{D}$, i.e. no charges are permuted nontrivially. We will now relax this assumption. Because the duality group is discrete, there are no nontrivial homomorphisms from the a continuous connected group to $\mathcal{D}$, and we can focus on discrete symmetries.
Consider a translation symmetry $G=\Z$. An obvious homomorphism from $\Z$ to $\mathcal{D}$ is to map the generator to $T^2$ (or $ST^{-2}S^{-1}$). To be concrete let us suppose
\begin{equation}
	T_{\hat{x}}: \quad  e\rightarrow e\, , \quad  m\rightarrow me^2 \, . 
	\label{eqn:tx}
\end{equation}
If this is the case, we expect that the magnetic monopole can not move in the $x$ direction since it changes its topological charge type under translation. In this case, the magnetic charges are fractonic even in the absence of any global symmetry, thus we call them intrinsically fractonic. We should stress that, while we assume translation invariance to facilitate the argument, it should be considered as a mathematical way to formulate the notion of mobility and the fractonic behavior does not rely on a precise translation symmetry of the Hamiltonian.

More generally, if $T_{\hat{x}}$ maps to any element of $\mathcal{D}$ which has infinite order, certain dyons become immobile (along $\hat{x}$).

\subsection{Gauged layered SPT phases}

We now present a general construction that realizes all kinds of fractonic $\U$ gauge theories discussed so far, thus providing a unified view of them. 

We start from a 3D SPT phase consisting of layers of 2D $\U_\text{g}\times G$ SPT phases, where $\U_\text{g}$ will become a gauge symmetry and $G$ is the global symmetry. Without loss of generality, let us assume that the layers are in the $yz$ plane. For simplicity we consider bosonic systems for now.
Such 2D SPT phases are classified by $\H^3[\U_\text{g}\times G, \U]=\Z\times \H^1[G, \U]$~\cite{chen2013}. Here the first $\Z$ factor corresponds to bosonic integer quantum Hall (BIQH) states~\cite{LuPRB2012, LevinPRL2013}, and the $\H^1[G, \U]$ factor describes the $G$ charge (i.e. one-dimensional representation) carried by a $2\pi$ $\U_\text{g}$ flux insertion. We also consider weak tunneling couplings between layers to allow charges to move in three dimensions.

Now we gauge the global $\U_\text{g}$ symmetry. We expect that the result is a deconfined 3D $\U$ gauge theory because before gauging the state is short-range entangled. The fundamental charged bosons become the $e$ particles, which can move freely in space, albeit with anisotropic dispersion. However, the layered SPT matter can affect the symmetry-enriched order in the resulting gauge theory. In particular, we know from previous studies of 3D topological insulating phases~\cite{MetlitskiPRB2013,wang2014, ZouPRB2018a, ZouPRB2018b} that symmetry properties of magnetic monopoles directly reflect the SPT order of the underlying matter.

\begin{figure}[t]
	\centering
	\includegraphics[width=0.8\columnwidth]{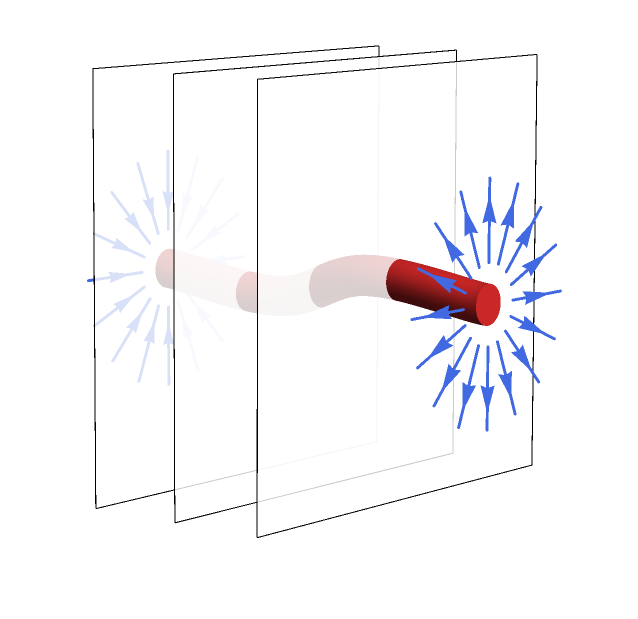}
	\caption{A monopole-antimonopole pair in a layered SPT phase. A Dirac string penetrating layers of 2D SPT states is illustrated.}
	\label{fig:monopole}
\end{figure}

We start the analysis from gauged layers of BIQH states. Suppose each layer has a Hall conductance ${\sigma_H=2}$. Physically, if we insert a $\U$ monopole $m$ in such a phase, a Dirac string of $2\pi$ flux is attached to the monopole, which penetrates the BIQH layers in the half space. Now suppose we have a monopole-antimonopole configuration. Due to the quantum Hall response, each $2\pi$ flux penetration contributes two units of electric charge, and thus the configuration has a total electric charge proportional to the distance between the monopole and the antimonopole. Thus the monopole in this layered SPT phase exhibits a kind of ASOC. This is illustrated in Fig. \ref{fig:monopole}.

Let us make this argument more precise. We start with a charge-neutral monopole, and translate it by one step along $\hat{x}$,  across one BIQH layer. Suppose this can be done using gauge-invariant (i.e. charge-neutral) local operators. This process can be viewed as an instanton tunneling event in the 2D BIQH state. Due to the quantum Hall effect, the $-2\pi$ flux acquires a polarization charge $-2$. In order to conserve charge, a local (gapped) boson excitation with $+2$ charge must also be created, in accordance with the transformation in Eq.~\eqref{eqn:tx}. This is typical for fractonic excitations: moving them requires the creation of additional excitations, and thus there is an energetic barrier. In the fully gauged theory, monopoles cannot move between layers in the $\hat{x}$ direction. We can see that this construction precisely realizes the symmetry transformation given in Eq. \eqref{eqn:tx}, and magnetic charges become intrinsically immobile (thus far in one direction only, but we may stack layers of BIQH states in three independent directions to make monopoles completely immobile fractons).

One question that we have not addressed thus far is the spectrum of the gauged layered SPT phase. Coupling to SPT matter generates Chern-Simons terms in each layer, which changes the photon spectrum. A very similar state was studied in \Ref{LevinFisher}, and we adapt their results to the present setting in Appendix~\ref{app:spec}. We find that the photon remains gapless, but its dispersion is softened in the direction perpendicular to the SPT layers.

Now we consider other phases classified by $\H^1[G, \U]$. Similarly, as the magnetic monopole is transported, the global $G$ charge is modified.  The magnetic monopole in the 3D state becomes fractonic if the $G$ charge has infinite order, for example when $G=\U$. In this case, each 2D layer has a ``crossed'' quantized Hall response between $\U_\text{g}$ and $\U$. This implies that as we separate a pair of monopoles, the Dirac string between them carries a growing $\U$ charge. This is essentially equivalent to the model discussed in Sec.~\ref{sec:U1gauge}, up to an S duality that swaps the electric and magnetic charges.

We also consider $\U$ gauge theories with fermionic matter (one can map a $\U$ gauge theory with fermionic matter to one with bosonic matter and a $\theta$ term with ${\theta=2\pi}$). In this case, the full dyon spectrum can be generated by a fermionic excitation $f$ with unit electric charge, and a neutral bosonic monopole $m$ with unit magnetic charge. Correspondingly, we can construct translation symmetry enriched $\U$ gauge theories by gauging layered fermionic SPT phases. We do not attempt a general classification here, opting instead to focus on two examples. 

First consider a 3D state made up of 2D layers given by fermionic integer quantum Hall (IQH) states with $\sigma_{H}=1$. Gauging the $\U$ symmetry turns the system into a $\U$ gauge theory with fermionic matter. The Hall response implies the following transformation of charge types under translation:
\begin{equation}
	T_{\hat{x}}: \quad  f\rightarrow f\, , \quad m\rightarrow mf \, ,
	\label{}
\end{equation}
which corresponds to the $T$ element in $\mathcal{D}$.

We can also take 2D $\U_\text{g}\times\U$ fermionic SPT phases as building blocks, characterized by crossed Hall response. We remark that although the crossed Hall conductance is the same as the bosonic case, the charge-1 excitations are fermions. After gauging, we find that the magnetic monopole $m$ becomes symmetry-enforced fractonic, while the fermionic charge $f$ remains free. If we instead label the charge lattice with a bosonic charge $e=m^\dag f$ and $m$ (corresponding to a T duality), both of them are fractonic. This is exactly the state we described in Sec. \ref{sec:classification}.

In this section we only considered translations along one direction. It is rather straightforward to generalize the construction to full three-dimensional translations, by stacking 2D SPT phases in all directions.

The construction also provides a new interpretation of the anomaly-vanishing condition derived in Sec. \ref{sec:classification}: if $\mb{v}_e$ and $\mb{v}_m$ were not parallel, all dyons become symmetry-enforced fractons. It is not clear how one could realize such a state from gauging layered 2D SPT phase. The anomaly-vanishing condition guarantees that such a situation does not arise, and one can realize any such symmetry-enforced fractonic matter using the gauged layers construction.

\section{Gauging fractonic symmetries}
	\label{sec:gauging}

	In this section we make a connection between our construction of fractonic models and more familiar constructions via higher-rank tensor gauge theory or gauging subsystem symmetries. We provide an explicit mapping between the gauged fractonic $\U$ gauge theory in Sec. \ref{sec:example} and a tensor gauge theory with $(1,1)$ scalar charge~\cite{PretkoPRB2017a, PretkoPRB2017b,BulmashPRB2018}. We also demonstrate that the $e$ particle in 3D toric code becomes fractonic when a subsystem symmetry is enforced.

	\subsection{Higher-rank tensor gauge theories}
For simplicity we work with a ``flat band'' version of a $\U$ gauge theory, where photons are dispersionless. This is achieved by dropping the $\mb{E}_{e}^2$ terms from the Hamiltonian in Eq.~\eqref{eqn:U1gauge}. While the dynamics is fine-tuned in such a limit, this does not affect the structure of the gauge-invariant Hilbert space that we are interested in. 

We have three independent $\U$ symmetries generated by shifting all rotors on $\hat{x},\hat{y}$ and $\hat{z}$ links respectively. 
To gauge these symmetries we introduce rotor variables representing new $\U$ gauge fields: three on each vertex labeled by $\hat{x},\hat{y},\hat{z}$, and a pair on each face, e.g. on an $\hat{x}$ face we introduce a $\hat{y}$ and $\hat{z}$ rotor. The rotor variables are denoted by $\tilde{E}$ and $\tilde{A}$, as before. Roughly speaking, the site gauge fields become identified with the diagonal components of a tensor gauge field, while the face gauge fields become the off-diagonal components.  As we shall see, the symmetric condition on the tensor gauge field is imposed energetically in our model.

Since the new gauge fields are sourced by electric fields of the original $\U$ gauge theory, for each edge we have a Gauss's law constraint 
\begin{align}
(\nabla_{\hat{i}} \cdot \mb{E})_{\vr, \hat{i}} :&= {E}_{\vr, \hat{i}}
+ {\tilde{E}}_{\vr+\mb{\hat{i}},\hat{i}} -  {\tilde{E}}_{\vr,\hat{i}}  + \sum_{p \ni ({\vr, \vr+\bf{\hat{i}}})} \epsilon_p^{\vr,\hat{i}} {\tilde{E}}_{p,\hat{i}}
\nonumber \\
&=0 
\, ,
\end{align}
where  $\epsilon_p^{\vr,\hat{i}}= \pm 1$ is $1$ if $p$ extends from $(\vr,\vr+\mb{\hat{i}})$ in a positive direction, and $-1$ otherwise. 

Additionally we follow the minimal coupling procedure to obtain
\begin{align}
\label{eq:mcterm}
- K \sum_{p} \cos [ (\nabla\times \mb{A})_{p} - {\tilde{A}}_{p,\hat{i}} + {\tilde{A}}_{p,\hat{j}} ]
\, ,
\end{align}
where $\hat{i}\neq\hat{j}$ are the axes parallel to $p$, chosen such that $\{ \hat{i},\hat{j},\hat{p} \}$ forms a right handed basis. 

We also introduce magnetic field terms to the Hamiltonian that penalize nonzero flux for $\tilde{\mb{A}}$:
\begin{align}
	- \lambda \sum_{\hat{i}} [ \sum_{c} \cos( \nabla_{\hat{i}} \times \tilde{\mb{A}} )_{c }  + \sum_{\vr, \hat{j}\neq \hat{i}} \cos( \nabla_{\hat{i}} \times \tilde{\mb{A}} )_{\vr,\hat{j}} ]
\, ,
\label{eqn:magnetic1}
\end{align}
where
\begin{align}
	( \nabla_{\hat{i}} \times \tilde{\mb{A}} )_{c } :&= \sum_{p \in c} \epsilon_{c}^{p,\hat{i}} {{\tilde{A}}}_{p, \hat{i}}
\, ,
\label{eqn:magnetic2}
\end{align}
and $ \epsilon_{c}^{p,\hat{i}}=+1$ if $(p,\hat{i})$ sits in a positive direction relative to the center of the cube $c$, $-1$ otherwise. 
Similarly,  
\begin{align}
	( \nabla_{\hat{i}} \times \tilde{\mb{A}} )_{\vr,\hat{j}} := {{\tilde{A}}}_{\vr+\mb{\hat{j}}, \hat{i}} - {{\tilde{A}}}_{\vr, \hat{i}} - \sum_{p \ni (\vr,\vr+\mb{\hat{j}})} \epsilon_{p}^{\vr,\hat{j}} {{\tilde{A}}}_{p, \hat{i}}
\, .
\end{align}

We would like to rewrite the model as a tensor gauge theory. To this end, let us perform the following transformation:
\begin{align}
\prod_{e}\,  \prod_{v \in e} CE_{e,(v,\hat{e})}^{\epsilon_e^v} \prod_{\Box \ni e} CE_{e,(\Box,\hat{e})}^{\epsilon_e^\Box}
\, ,
\end{align}
where $\epsilon_e^v$ is $\dagger$ when $v$ sits on the positive end of $e$ and $1$ otherwise, similarly $\epsilon_e^\Box$ is $\dagger$ when $\Box$ extends from $e$ in the positive direction and $1$ otherwise. The $CE$ gate acting on two rotors is defined by 
\begin{align}
CE_{1,2} \ket{A_1,A_2} &= \ket{A_1,A_1+ A_2}
\, ,
\\
CE_{1,2} \ket{E_1,E_2} &= \ket{E_1-E_2,E_2}
\, .
\end{align}
The above transformation can be viewed as a change of variables.

After this change of variables the newly introduced Gauss's law constraints become 
\begin{equation}
	(\nabla_{\hat{i}} \cdot \mb{E})_{\vr, {\hat{i}}}=0\mapsto {E}_{\vr,{\hat{i}}}=0 \, ,
	\label{}
\end{equation}
effectively fixing out the original rotor variables. 
Gauss's law of the original gauge fields $(\nabla \cdot \mb{E})_{\vr}$ in Eq.~\eqref{eqn:U1gauge} maps to 
\begin{align}
 \sum_{\hat{i}} - {\tilde{E}}_{\vr+\mb{\hat{i}},\hat{i}} + 2 {\tilde{E}}_{\vr,\hat{i}} - {\tilde{E}}_{\vr-\mb{\hat{i}},\hat{i}}
+\sum_{p \ni \vr}  \epsilon_p^{\vr}  \left({\tilde{E}}_{p,\hat{i}} +	 {\tilde{E}}_{p, \hat{j}} \right) 
\, ,
\end{align}
where we have imposed the new Gauss's law strictly, by setting $\tilde{E}_{\vr, {\hat{i}}}=0$.

We also find that the term in Eq.~\eqref{eq:mcterm} becomes 
\begin{align}
	-K \sum_p \cos (\tilde{{A}}_{p,\hat{i}} -\tilde{{A}}_{p,\hat{j}} )
\, ,
\end{align}
which gives an energy penalty to any state with fields on a face that are not symmetric. After projecting into the zero-energy subspace of this term, we have a symmetric tensor gauge field with a single independent rotor on each face defined by 
\begin{align}
{\tilde{A}}_{p,\hat{i}} \sim {\tilde{A}}_{p,\hat{j}} \mapsto {\tilde{A}}_{p}
\, ,
&&
\frac{1}{2}( {\tilde{E}}_{p,\hat{i}} + {\tilde{E}}_{p,\hat{j}} )  \mapsto  {\tilde{E}}_p
\, .
\end{align}
We remark that the factor of $1/2$ above leads to the unusual commutation commutation relation
\begin{align}
[ \tilde{A}_p, \tilde{E}_p] = \frac{i}{2}
\, ,
\end{align}
which was chosen to match the convention of Ref.~\onlinecite{BulmashPRB2018}. 

Within the symmetric subspace, 
the original Gauss's law becomes
\begin{align}
 \sum_{\hat{i}} - {\tilde{E}}_{\vr+\mb{\hat{i}},\hat{i}} + 2 {\tilde{E}}_{\vr,\hat{i}} - {\tilde{E}}_{\vr-\mb{\hat{i}},\hat{i}}
+ 2 \sum_{p \ni \vr}  \epsilon_p^{\vr}  {\tilde{E}}_{p} 
\, ,
\end{align}
 as expected for the symmetric tensor gauge theory with $(1,1)$ scalar charge. 
 Additionally the magnetic field terms in Eqs. \eqref{eqn:magnetic1} and \eqref{eqn:magnetic2} become 
 \begin{align}
 &\sum_{p \in c} \epsilon_{c}^{p,\hat{i}} {{\tilde{A}}}_{p} 
\, , 
\\
&{{\tilde{A}}}_{\vr+\mb{\hat{j}}, \hat{i}} - {{\tilde{A}}}_{\vr, \hat{i}} - \sum_{p \ni (\vr,{\hat{j}})} \epsilon_{p}^{\vr,\hat{j}} {{\tilde{A}}}_{p} 
\, ,
\end{align}
matching those of the symmetric tensor gauge theory with $(1,1)$ scalar charge. 

\subsection{Type-I fracton models}
We now discuss connections to type-I fracton spin models, particularly the X-cube model~\cite{VijayPRB2016}. We point out that $\U$ having an infinite number of irreducible representations, indexed by integers, was essential in our constructions above.  Another class of examples with infinite symmetry groups are systems with subsystem symmetries~\cite{VijayPRB2016,WilliamsonPRB2016,you2018subsystem,YYZ2018,Devakul2018b,devakul2018universal,subsystemphaserel,ShirleyGauging2018,kubica2018ungauging,Williamson2018}, which also possess an infinite number of irreducible representations in the thermodynamic limit. Let us consider a concrete example, namely a 3D $\Z_2$ toric code with planar subsystem symmetries. We define the toric code following the standard convention, with qubits on edges of a cubic lattice: 
\begin{equation}
	H=-\sum_\vr \prod_{\vr'\in \text{NN}(\vr)}X_{\vr\vr'} - \sum_{\Box} \prod_{e\in \Box}Z_e.
	\label{eqn:3dtc}
\end{equation}
There are two types of excitations: violations of the vertex terms are particle excitations, denoted by $e$, and violations of plaquette terms are loop excitations.

The Hamiltonian respects a spin-flip subsystem symmetry, generated by $X^{\otimes L \times L}$, on each dual plane. For example, on an $xy$ plane we have
\begin{equation}
	\prod_{x,y}X_{(x,y,z), \hat{z}} \, .
	\label{}
\end{equation}
Under this subsystem symmetry the $e$ particle becomes fractonic, as the associated string operators are products of $Z$'s along lattice paths and hence any motion of $e$ changes one of the subsystem symmetry charges. We notice that such a symmetry can actually be defined on any closed surface on the dual lattice, which is an example of a ``1-form'' symmetry in the $\Z_2$ gauge theory~\cite{Gaiotto2015}, however we only make use of a rigid subset defined on dual planes.

We can then gauge all the subsystem symmetries~\cite{VijayPRB2016, YYZ2018, ShirleyGauging2018}. 
To do so we introduce two gauge qubits onto every face, one for each dual plane crossing the face. For example we associate qubits $(\Box,\hat{x}),\,(\Box,\hat{z})$ to a face $\Box$ in a $\hat{y}$ plane. 
These gauge variables come with a Gauss's law constraint on each edge, e.g. 
\begin{align}
X_{\vr,\hat{z}}  \prod_{\Box \ni (\vr,\hat{z})} {X}_{\Box,\hat{z}}
\, .
\end{align}
We also introduce Hamiltonian terms for every cube and dual plane that penalize nonzero gauge flux on the faces of the cube that intersect the dual plane e.g. 
\begin{align}
- \prod_{(\Box,\hat{z})\in c} Z_{\Box,\hat{z}}
\, ,
\end{align}
and similar terms for $\hat{x}$ and $\hat{y}$. In the above $c$ denotes a single cube in the lattice, note the product involves four $Z$ matrices. 
The $Z$ terms in the Hamiltonian are modified by minimal coupling as 
\begin{align}
- Z_{\Box,\hat{x}} Z_{\Box,\hat{z}}
\prod_{e\in \Box}Z_e 
\, ,
\label{eqn:tcZ}
\end{align}
for a face in the $\hat{y}$ plane, and similarly for $\hat{x},\hat{z}$.  

It is possible to disentangle the original matter qubits and gauge constraints from the newly introduced gauge qubits by applying a circuit of controlled-$X$ gates~\cite{Gaugingpaper,williamson2014matrix,NewSETPaper2017} 
\begin{align}
\label{eq:cxcircuit}
\prod_{e}
\prod_{\Box \ni e}
CX_{e,(\Box,\hat{e})}
\, ,
\end{align}
where
\begin{align}
CX_{1,2} \ket{i,j} := \ket{i,i+j}
\, ,
\end{align}
in the $Z$ basis. The $CX_{1,2}$ gate commutes with $Z_1$ and $X_2$, while
\begin{align}
CX_{1,2} X_1 CX_{1,2}^{\dagger} &= X_1 X_2
\, ,
\\
CX_{1,2} Z_2 CX_{1,2}^{\dagger} &= Z_1 Z_2
\, .
\end{align}
Under the transformation in Eq.~\eqref{eq:cxcircuit} the Gauss's law constraints are mapped to $X_{e}$. We focus on the sector where these constraints are all satisfied,  and hence all edge qubits are projected into the ${\ket{+}:=(\ket{0}+\ket{1})/\sqrt{2}}$ state. The $Z$ term of the Hamiltonian Eq. \eqref{eqn:tcZ} is mapped to 
\begin{align}
- Z_{\Box,\hat{x}} Z_{\Box,\hat{z}}
\, ,
\end{align}
for a $\hat{y}$ plaquette, and similarly for $\hat{x}$, $\hat{z}$. 
The $X$ term in Eq. \eqref{eqn:3dtc} is mapped to 
\begin{align}
-  \prod_{\Box \ni \vr} X_{\Box,\hat{i}} X_{\Box,\hat{j}}
\, ,
\end{align}
where $\hat{i} \neq \hat{j}$ label the two qubits on each face. The flux term for the new gauge fields in the Hamiltonian remains unchanged. 

Next we restrict to the sector where the $Z_{\Box,\hat{i}} Z_{\Box,\hat{j}}$ terms are all satisfied, this leaves a single qubit degree of freedom on each face defined by new operators: 
\begin{align}
{Z_{\Box,\hat{i}}\sim Z_{\Box,\hat{j}} \mapsto Z_\Box}\, , && 
X_{\Box,\hat{i}}X_{\Box,\hat{j}} \mapsto X_\Box \, . 
\end{align}
The Hamiltonian then becomes 
\begin{align}
H_{\text{gauged}} = 
-\sum_{\vr} \prod_{\Box \ni \vr} X_\Box -  \sum_{c,\hat{i}} \prod_{\hat{i} \in \Box \in c} Z_{\Box}
\, ,
\end{align}
which is precisely the X-cube Hamiltonian on the dual cubic lattice. The cube term can be traced back to the site stabilizer in the (ungauged) toric code Hamiltonian. Thus the fracton in the X-cube model is indeed the gauge charge of the toric code model as expected.

\section{Discussion}
To conclude the paper we outline some open questions and future directions: 
In the classification of $\U$ gauge theory enriched by translation symmetry, we only considered a simple class of charge permutations, namely those generated by $T$. There are (infinitely) many other types of permutation that are not conjugate to those generated by $T$. It would be interesting to find out whether such symmetry actions could be realized physically. In the most general classification, one must consider twisted group cohomology corresponding to fractionalization on charge excitations.

It would be interesting to generalize our construction to non-Abelian gauge theories. It is well-known that in $(3+1)$D pure non-Abelian gauge theories are in the confined phase. They can become deconfined by coupling to gapless matter. For example, an $\mathrm{SU}(N_c)$ gauge theory coupled to $N_f$ flavors of massless Dirac fermions in the fundamental representation flows to the free fixed point when $N_f\gg N_c$ and realizes a non-Abelian Coulomb phase. We may add gapped matter to the theory without affecting the RG flow. This raises the possibility of constructing examples of fractonic non-Abelian gauge theories in the presence of both gapless and gapped matter. Another direction would be to generalize the gauged 3D stacked quantum Hall states to non-Abelian symmetry, which could alter the infrared properties of the gauge theory.

We discussed briefly how a $\Z_2$ gauge theory in 3D can be enriched by planar subsystem symmetries, the gauging of which produces the X-cube model (in a certain sector). We believe this example can also be described using the symmetry fractionalization formalism, where the fractonic behavior of $e$ charges corresponds to a certain cohomology class in $\H^2[\Z_2^\text{sub}\rtimes\Z, \Z_2]$, where $\Z_2^\text{sub}$ is the (extensively large) group of planar subsystem symmetries, and $\Z$ is the translation symmetry group. Similarly, one can classify symmetry fractionalization on loop excitations by $\H^3[\Z_2^\text{sub}\rtimes\Z, \Z_2]$. 
We leave the investigation of properties of such loop excitations to future work. 
 It would also be worthwhile understanding the relation between the 't Hooft anomalies for global $U(1)\times \Z^3$ we have studied and those of a higher form or subsystem $U(1)$ times translation symmetry~\cite{Kobayashi2018}.

It will be interesting to see whether similar ideas can be useful for type-II fractons with fractal dynamics, such as those in Haah's cubic code~\cite{Haah}, or analogous $\U$ models~\cite{HaahTalk,bulmash2018generalized}.

\vspace{.5cm}
\emph{Related work:} Recently, several relevant works on symmetry-enforced fractonic matter appeared~\cite{Pretko2018,PretkoGauge2018,KumarArxiv2018}.

\section{Acknowledgement}
MC thanks Xie Chen and Yizhi You for enlightening conversations. DW thanks David Aasen and Kevin Slagle for useful discussions. 
The work is supported by start-up funds at Yale University (DW and MC) and NSF CAREER under award number DMR-1846109 (MC). ZB acknowledges
support from the Pappalardo fellowship at MIT.

\appendix

\section{'t Hooft Anomaly in $\U$ gauge theory} 
\label{app:anomaly}
In this Appendix we consider a $\U$ gauge theory with a global symmetry group $G=\U\times\Z^3$, and assume that symmetry transformations in $G$ do not permute charge types. $G$ symmetry-enriched $\U$ gauge theories are classified by the projective representations carried by electric and magnetic charges, denoted by $[\omega_e]$ and $[\omega_m]$ respectively, both of which are in $\H^2[G, \U]$. 
For convenience we represent $\U$ as $\mathbb{R}/\Z$, with multiplication denoted additively. We also define $[n_m]\in \H^3[G, \Z]$, which is given by $n_m=\delta \omega_m$ where $\delta$ is the boundary operator.

The 't Hooft anomaly associated with $G$ is characterized by the following group $5$-cocycle in $\H^5[G, \U]$~\cite{NingU1}:
\begin{equation}
	O_5(\mb{g,h,k,l,m})=e^{2\pi i\omega_e(\mb{g,h})n_m(\mb{k,l,m})} \, .
	\label{}
\end{equation}
Physically, if $O_5$ is cohomologically nontrivial, the corresponding symmetry-enriched $\U$ gauge theory must live on the boundary of a (4+1)d bosonic SPT phase described by $O_5$.

Before calculating $O_5$ we need to understand possible (4+1)d bosonic SPT phases with $\U\times \Z^3$ symmetry. We could obtain the classification by directly computing the group cohomology~\cite{ChengPRX2016}, but we instead take a more intuitive approach.  Since the elements of $\Z^3$ are actually translations, such SPT phases can be represented using a layer construction, i.e. as stacks of lower-dimensional SPTs with only $\U$ symmetry. We remark that nontrivial $\U$ SPT phases only exist in even spatial dimensions. In this case, the only relevant option is to stack 2D $\U$ SPT planes, along any two of the directions (the other options is a 4D lattice of ``points'', or 0D $\U$ SPT states, however this state requires the full 4D translation $\Z^4$).

In \Ref{ChengPRX2016} it was shown that the stacking corresponds mathematically to the slant product for group cocycles. Therefore, to extract the cohomology class, we can apply the slant product to $O_5$ with respect to two translations, and define $O_3^{\mb{n}_1,\mb{n}_2}=i_{T_{\mb{n}_1}}i_{T_{\mb{n}_2}}O_5$, and then restrict $O_3^{\mb{n}_1,\mb{n}_2}$ to the $\U$ subgroup of $G$. The result must be a $3$-cocycle classified by $\H^3[\U, \U]=\Z$. So we can parametrize
\begin{equation}
	O_3^{\mb{n}_1,\mb{n}_2}({\theta}_1, {\theta}_2, {\theta}_3)=K_{\mb{n}_1,\mb{n}_2}\theta_{1}\frac{\theta_{2}+\theta_{3}-[\theta_{2}+\theta_{3}]_{2\pi}}{2\pi} \, .
	\label{}
\end{equation}

We also need explicit representatives of 2-cocycles in $\H^2[\U\times \Z^3, \mathbb{R}/\Z]$. We only consider those that correspond to ASOC. Denote elements of $\U\times\Z^3$ by $(\theta, \mb{n})$ where $\theta\in [0,2\pi), \mb{n}\in\Z^3$. We have
	\begin{align}
		\omega( (\theta_1,\mb{n}_1), (\theta_2, \mb{n}_2)) = \frac{\theta_1\mb{v}_e\cdot\mb{n}_2}{2\pi}\, , \qquad  \mb{v}_e\in\Z^3 \, .
		\label{}
	\end{align}

	The fact that this is a nontrivial cocycle for any $\mb{v}\neq 0$ follows from ${\omega( (\theta,0), (0, n))-\omega((0,n), (\theta,0))=\frac{\theta \mb{v}\cdot\mb{n}}{2\pi}}$.

	A direct calculation finds 
	\begin{equation}
		n( (\theta_1, \mb{n}_1), (\theta_2,\mb{n}_2), (\theta_3, \mb{n}_3))
		=\frac{\theta_{1}+\theta_{2}-[\theta_{1}+\theta_{2}]_{2\pi}}{2\pi}\mb{v}_m\cdot \mb{n}_{3}\, .
		\label{}
	\end{equation}

	After taking slant product (with respect to translations along $\mb{n}_1$ and $\mb{n}_2$) and restricting to the $\U$ subgroup, we get
	\begin{equation}
		K_{\mb{n}_1,\mb{n}_2}=(\mb{v}_m\cdot\mb{n}_1)(\mb{v}_e\cdot\mb{n}_2) - (\mb{v}_e\cdot\mb{n}_1)(\mb{v}_m\cdot\mb{n}_2).
		\label{}
	\end{equation}
	As expected we have $K_{\mb{n}_1,\mb{n}_2}=-K_{\mb{n}_2,\mb{n}_1}$.
	Considering all $\mb{n}_1,\mb{n}_2$, the condition can be summarized as 
	\begin{equation}
		\mb{v}_e\times \mb{v}_m=0.
		\label{}
	\end{equation}

	We can generalize the above calculation to multiple $\U$ symmetries. We denote the group elements of $\U\times \cdots\U$ by a vector $\bm{\theta}=(\theta_1, \theta_2, \dots)$, and parametrize $3$-cocycles in $\U\times \cdots \U$ by 
\begin{equation}
	O_3^{\mb{n}_1,\mb{n}_2}(\bm{\theta}_1, \bm{\theta}_2, \bm{\theta}_3)=\sum_{a b}K_{ab}\theta_{1a}\frac{\theta_{2b}+\theta_{3b}-[\theta_{2b}+\theta_{3b}]_{2\pi}}{2\pi}\, .
	\label{}
\end{equation}
Here one should notice that the 3-cocycles corresponding to $K_{ab}$ and $K_{ba}$ for $a\neq b$ are in fact equivalent.

The expressions given above for $2$- and $3$-cocycles of fractionalization classes are easily generalized:
		\begin{equation}
			\begin{gathered}
			\omega_e( (\bm{\theta}_1,\mb{n}_1), (\bm{\theta}_2, \mb{n}_2)) =\sum_{a,\alpha} \frac{v_e^{a\alpha}\theta_{1a}n_{2\alpha}}{2\pi}\, ,
			\\
			\begin{split}
		n_m( &(\theta_1, \mb{n}_1), (\theta_2,\mb{n}_2), (\theta_3, \mb{n}_3)) \\ 
		&=\sum_{a,\alpha} v_{m}^{a\alpha}\frac{\theta_{1a}+\theta_{2a}-[\theta_{1a}+\theta_{2a}]_{2\pi}}{2\pi}{n}_{3\alpha}
		\, .
	\end{split}
			\end{gathered}
		\label{}
	\end{equation}

	A straightforward calculation yields
	\begin{equation}
		K_{ab}=\sum_{\alpha\beta}(v_e^{b\alpha}v_m^{a\beta} - v_e^{b\beta}v_m^{a\alpha})n_{1\alpha}n_{2\beta} \, .
		\label{}
	\end{equation}
	The anomaly vanishes when $K_{ab}+K_{ba}=0$.

	\section{Spectrum of gauged layered SPT phases}
	\label{app:spec}
	In this Appendix we study the bulk spectrum of gauged layers of 2D SPT phases. We start from the simple case of layers of 2D quantum Hall states stacked along the $\hat{z}$ direction. We describe the low-energy physics using the following effective gauge theory after integrating out the matter:
	\begin{equation}
		\begin{split}
			\mathcal{L}=& \sum_z\frac{n}{4\pi}\varepsilon^{\mu\nu\lambda}A_{z,\mu}\partial_\nu A_{z,\lambda}\\
		&+ \sum_z\frac{1}{2g_3a}(\partial_0 A_{z,3}-A_{z,0}+A_{z+1,0})^2\\
		&- \sum_i J_3a(\partial_i A_{z,3}-A_{z,i}+A_{z+1,i})^2
		\, ,
		\end{split}
		\label{}
	\end{equation}
	 where $A_{z\mu}(x_0, x_1,x_2)$ is the lattice gauge field, $x_0$ is time, $x_1,x_2$ are coordinates in the $xy$ plane, $z$ is the layer index, $a$ is the lattice spacing, and $g_3$ and $J_3$ are coupling constants. We take $n=1$ for fermionic IQH and $n=2$ for bosonic IQH states. We have ignored terms that are irrelevant at low energies. 

	Let us try to understand the gapless gauge boson excitations in greater detail. One can find the dispersion relation for these gapless modes by going to Fourier space. For small momentum $\mb{k}$, we have
	\begin{equation}
		\begin{split}
		\mathcal{L}=&\frac{n}{4\pi a}\varepsilon^{\mu\nu\lambda}A_{\mu}ik_\nu A_{\lambda}\\
		&+\frac{1}{2g_3a} (k_0A_3-k_3A_0)^2 - \frac{J_3a}{2}\sum_i ( k_i A_3-k_3A_i)^2 \, .
		\end{split}
		\label{}
	\end{equation}
	The theory can now be diagonalized in the temporal gauge $A_0=0$. Following \Ref{LevinFisher} we obtain the dispersion
	\begin{equation}
		\omega^2=J_3g_3a^2(k_1^2+k_2^2)+\frac{4\pi^2 J_3^2a^4}{n^2}k_3^4.
		\label{}
	\end{equation}
	Hence the gauge bosons``soften'' along the $z$ direction.

\vspace{.5cm}
\bibliography{fracton}
\end{document}